\documentclass{article}
\usepackage{graphicx} 
\usepackage{hyperref}
\usepackage{indentfirst}

\title{Best practices for obtaining astrometric observations from JWST (274) NIRCam data}
\author{Marco Micheli, Bryan J. Holler}
\date{2026 March 16}

\begin{document}

\maketitle

\section{Introduction}

The James Webb Space Telescope (JWST), with its near-infrared imager NIRCam, is capable of obtaining detections of small bodies down to what would correspond to a visual magnitude of about 31. However, the complexity of NIRCam introduces multiple subtle effects in the way the images are obtained, and they in turn affect the astrometric information that can be extracted from the images. In the following, we very briefly summarize these peculiarities, and provide some basic information on how to extract meaningful astrometric positions from NIRCam images. This guide only touches the most basic and general points of NIRCam astrometry: for a detailed understanding of each specific dataset, we refer the user to the STScI documentation, and recommend direct interaction with their staff, who can usually provide extremely helpful information essential to understand a dataset.

\section{Astrometric uncertainties}

\subsection{When using WCS solution directly} \label{wcs}

The first important point is that, unfortunately, the WCS solution included in the images by the JWST pipeline is known to be biased, at the level of a few tenths of an arcsecond. This immediately implies that any astrometry extracted without solving the images cannot reach the intrinsic accuracy level of which JWST is capable.\\

\textbf{Recommendation:} Any time an image is not solved with field stars, it is recommended to assume a conservative astrometric uncertainty (rmsRA and rmsDec in ADES) of ±0.3$''$ at least. \\

The uncertainty shall be further increased in case of multiple measurements from the same telescope pointing, since the bias is observed to be a systematic effect, and therefore all measurements in the same sequence need to be considered correlated with each other. It is recommended to increase the uncertainty of each measurement by at least $\sqrt{N}$, where $N$ is the number of measurements sharing the same spacecraft pointing.

\subsection{On-plate solving}

The gold standard for JWST astrometry (and astrometry in general) is to solve each individual image with Gaia field stars directly detected in the image itself. Only Gaia stars present in catalog versions DR2 and later, for which proper motions are available, can be used for high-precision astrometric purposes. It is also important to ensure that all stars used in the solution actually have proper motions included in the catalog, and are properly corrected using them to the epoch of the observation.\\

NIRCam operates 2 separate wavelength channels. The long wavelength channel has a monolithic detector covering a field of view of 2.2$'$×2.2$'$. The short wavelength channel is composed of a mosaic of 4 detectors, each covering a 1.1$'$×1.1$'$ field of view, and separated by gaps of approximately 5$''$.\\


Even in a 2.2$'$×2.2$'$ FoV, low-density areas of the sky may occasionally not contain a number of Gaia stars sufficient for a full astrometric solution if the object is transiting through a region of sky far from the galactic plane, where field stars are scarce.\\

The scarcity of reference stars is often more severe than it might seem, because Gaia is an optical catalog, while NIRCam images are in the near infrared: some Gaia stars with a mostly blue color may therefore not be visible at all in the JWST exposures, and therefore become unusable for the astrometric solution.\\

Similarly, some stars may be too bright and therefore fully saturated in NIRCam images. Other stars, as mentioned above, might lack proper motion, and therefore be unsuitable for a high-precision solution.

In summary, extreme care should always be placed when scheduling an observation with astrometric purposes, in order to ensure that a sufficient number of usable reference stars are present in the FoV.\\

\subsubsection{Processing level} \label{solution}

In order to obtain a proper astrometric solution of a NIRCam image, it is important to understand the different processing stages provided by the JWST pipeline. Some of the processing steps are astrometrically important, and should be preserved, while some other steps, introduced for non-astrometric purposes, have a detrimental effect on the images, and need to be skipped to obtain images that can be astrometrically solved.\\

The most important valuable contribution provided by the JWST pipeline is the rectification of the images. As mentioned above, NIRCam's images often contain a very limited number of reference stars, and it is therefore impossible to constrain a solution at any order higher than linear. Fortunately, JWST's pipeline provides distortion-corrected images (the so-called I2D processing level, one of the products available directly on MAST), and these have been shown to be good at a level that is sufficient to assume a linear distortion model for the entire FoV. Official JWST documentation \footnote{\url{https://jwst-docs.stsci.edu/jwst-calibration-status/nircam-calibration-status/nircam-imaging-calibration-status}} states that the distortion corrections should be good to the level of a few milliarcseconds, and are therefore not the limiting factor of a typical astrometric measurement.\\

Unfortunately, a later step of the standard JWST pipeline processing is detrimental for astrometry. JWST's pipeline is designed to process stationary objects, and makes use of the assumption that real astronomical sources remain in a stable position during each integration. This assumption is heavily used to recognize image artifacts, such as cosmic rays, and remove them from the final product: any pixel that experiences an abrupt change of flux during the exposure is assumed to have been involved with an anomaly, and the jump is artificially removed during processing. However, when a moving object is tracked with non-sidereal motions, field stars trail in the frame, and therefore a given pixel may see a star entering and/or exiting it during the exposure. This results in an abrupt flux jump (when the star crosses the pixel), and is recognized as a likely artifact by the pipeline, which applies its corrections and tries to remove the source. The result is that most field stars are corrupted by this procedure, with their bright cores being removed, leaving only the fainter wings visible. In order to avoid this issue, it is necessary to specifically disable the corresponding step in the JWST data processing pipeline (the so-called NOJUMP mode), ensuring that field stars are fully preserved and remain visible in the images, and usable as astrometric references. This change of course comes with the side effect that cosmic rays are no longer removed, and may occasionally affect the target or the reference stars.\\

In addition to the NOJUMP custom processing, images to be used for astrometric solving also benefit from an additional tweak in the processing pipeline. The default configuration attempts to recover flux information on saturated stars with an extrapolation procedure. This algorithm often creates processing artifacts near the core of bright stars, which affect the ability to use them as astrometric references. This saturation compensation mode can be turned off, resulting in images (the so-called NOSAT mode) that are significantly better behaved for astrometric use, at the cost of invalidating flux preservation.\\

The use of NOJUMP+NOSAT images, always with I2D distortions applied, is therefore recommended for astrometric purposes. Unfortunately, NOJUMP+NOSAT images are not default products of the pipeline, and they need to be produced by running the pipeline locally. See Appendix \ref{appendix} for details.\\

These images usually display clean and uncorrupted star trails, and can be used to extract a linear astrometric solution of the field. The target itself, on the other hand, will be unaffected by these processing artifacts introduced by the JUMP and SAT corrections if JWST is tracking non-sidereally on the target: it is therefore recommended to measure it on the original I2D frames, which benefit from the JUMP processing and are usually less contaminated by cosmic rays.\\

There is one final important caveat to be considered when using non-sidereally tracked NIRCam images for astrometry. NIRCam's readout is complex, and it happens over a timescale of about 10 seconds, row by row. This has crucial implications for timing, which will be discussed in the next section. However, it also has implications for the astrometric solution itself, since it introduces a linear stretch of the image in the tracking direction. This is easily captured by a proper linear astrometric solution, but it should nevertheless be noted, because it results in a significantly different plate scale in the two directions.\\

\textbf{Recommendation:} When observing high-profile objects, where mas-level astrometry is needed, images must be astrometrically solved with Gaia reference stars. A linear solution is sufficient, if using images rectified by the JWST pipeline. It is recommended to use regular I2D images to extract the location of the target, but switch to NOJUMP+NOSAT processing products to extract the centroid of the stellar trails to be used for the astrometric solution. The two can then be combined to produce the final astrometry.

\subsubsection{Partial solutions}

The procedure outlined in Sect. \ref{solution} is the gold standard to solve an image for astrometric purposes. However, it relies on the existence of a sufficient number of usable Gaia stars in the field. There are situations when that condition may be impossible to meet. When this is the case, it is possible to rely on further assumptions to ensure a meaningful astrometric measurement can be obtained with fewer reference stars.\\

The basic idea of this intermediate approach is to still make use of the provided WCS frame, but only for aspects that can be assumed to be correct at the level of precision we require. Within these constraints, we can assume that the provided WCS solution is correct in scale and rotation, and only solve for its translational shift with respect to the actual sky, which is dependent on the absolute pointing of JWST and can be off by ±0.3$''$ (see \ref{wcs}).\\

We can use a smaller number of Gaia stars in the field to only estimate for this offset. Ideally, even just one star would be enough, but in order to ensure some statistics, and estimate the uncertainties of the process, it is important to have at least a few sources, and use the RMS of the biases as an estimate for the errors introduced by this simplified procedure in the astrometric solution determination.\\

In addition to this simplification, we can also assume that whatever bias between the embedded WCS and reality is caused by the spacecraft pointing, and is therefore identical across the various chips. If necessary, we can therefore solve for the offset using stars in all chips, increasing the number of Gaia references for the calibration. \\

\textbf{Recommendation:} If the number of Gaia stars is insufficient to fully solve the chip, it is reasonable to assume the provided WCS is correct in all parameters except for a translational shift along both axes. This shift can be determined using fewer reference stars, even across chips if necessary. The RMS of the determined offsets over all stars can then be used as a conservative measure of the astrometric solution's uncertainty.

\section{Timing uncertainties}

The previous section provides an overview of the most critical steps necessary to extract accurate positional measurements from a NIRCam frame. Astrometry of small bodies is, however, not just a measurement of position, but also of an associated time corresponding to each position.\\

This step is unfortunately even more complex for NIRCam, due to the peculiar readout process used for its detector. Contrary to optical CCDs and CMOSes, infrared imaging arrays are usually read out continuously, without clearing the image, with the so-called non-destructive up-the-ramp readout process \footnote{\url{https://jwst-docs.stsci.edu/jwst-near-infrared-camera/nircam-instrumentation/nircam-detector-overview/nircam-detector-readout-patterns}}. Each readout happens sequentially across the detector, and the timescale of the sweep from top to bottom is not negligible for astrometric purposes, since it covers a time window of 10.737 s. Consequently, each row of the detector will have timetags that might differ by as much as 10 s, not negligible for astrometric purposes on all but the slowest targets.\\

\textbf{Recommendation:} if no readout-specific timing corrections are applied, it is recommended to assume a conservative time bias uncertainty (uncTime in ADES) of 10 s at least. \\

Since it is likely that an object would always fall on approximately the same area of the same detector, the above time uncertainty should be assumed as a systematic bias.

\subsection{Row-level timing determination}

When sub-second timing accuracies are needed, in addition to the readout process outlined above, it is important to note that the readout proceeds in different directions depending on which detector is used \footnote{\url{https://jwst-docs.stsci.edu/jwst-near-infrared-camera/nircam-instrumentation/nircam-detector-overview/nircam-detector-readout}}. Therefore, particular care needs to be placed to ascertain which specific detector is being used for the image.\\

Furthermore, each readout mode is characterized by a different number and relative position of skipped frames and reset intervals, including for example a reset cycle at the beginning of each integration\footnote{\url{https://jwst-docs.stsci.edu/accessing-jwst-data/jwst-science-data-overview/jwst-time-definitions}}. Each of these gaps also happen on an effective time scale of 10.737 s, and they need to be properly understood and taken into account to avoid introducing further timing biases at the 10 s level.\\

For a full understanding of the process, we refer the reader to JWST's documentation\footnote{\url{https://jwst-docs.stsci.edu/understanding-exposure-times}}, and encourage a thorough understanding of the specific readout pattern used for the observation. Once the proper readout pattern, direction and line placement of the source is taken into account, it is possible to determine the exact time tag of a detection to a level of about 0.1 s.\\

Please note that reaching this level of timing accuracy, or better, also requires taking the distortion correction into account: the exact row line of a certain source shall be determined on the original images (ideally, the CAL processing level), not on the ones that have been distortion-corrected for astrometric purposes (the I2D level).\\

\textbf{Recommendation:} when observing fast-moving objects, such as NEOs, where sub-second timing precision is needed, it is necessary to consider the exact readout mode used for the specific dataset to achieve the required timing precision. When these corrections are performed, it is possible to reach the effective limits of the instrument, which can be conservatively estimated at around 0.1 s (for both the systematic component, uncTime, and the random component, rmsTime, in ADES).

\section{Spacecraft location}

The final ingredient of each astrometric measurement is the indication of the location of the observer at the time of the observation. JWST is a spacecraft, and it is therefore in constant motion with respect to the Earth. The ADES format requires the submitter to specify the coordinates of the spacecraft at the time of the observation.\\

The process of retrieving spacecraft coordinates is common to all spacecraft, and we won't discuss it in detail in this guide. We would nevertheless point out a few important points that need to be taken into account when determining the coordinates.\\

The first point is associated again with the timing uncertainties. The spacecraft location is associated to a given time, and the discussion in the previous section highlights how easy it is to introduce 10 s level time biases. For a moving spacecraft, this bias will also indirectly affect our knowledge of the observer location, introducing a second-order effect that needs to be considered in the orbit determination process.\\

For high precision purposes, when an accurate location of the spacecraft is needed, we also recommend to retrieve the spacecraft coordinates at least a couple of weeks after the observations are acquired. Most spacecraft ephemeris providers receive the most accurate reconstructed trajectory only a posteriori, once sufficient tracking data has been retrieved from the spacecraft. If astrometry is reported right after the observations, the associated location might be based on preliminary or predicted ephemerides, which might be slightly off from the real a posteriori reconstruction of the spacecraft motion.\\

One final reminder is to pay attention to the appropriate timescale and reference system of the associated ephemeris. ADES requires positions at a specific time in UTC, and referred to the ICRF reference frame. Some ephemeris providers may either use different timescales (e.g., TDB) or reference systems (e.g., EME2000), and if so the values need to be corrected accordingly.\\

\textbf{Recommendation:} spacecraft coordinates need to be included in the submitted astrometric record (pos1, pos2 and pos3 in ADES, plus the appropriate sys and ctr keywords for the reference system). For high-precision use, it is necessary to pay particular attention to the quality of the spacecraft ephemeris, and to the reference systems used in the report.\\

It is also recommended, although not strictly required, to include the velocity components of the spacecraft in the same reference system. Velocity components are necessary to properly account for timing uncertainties in the orbit determination process.

\section{Conclusions}

This guide only outlines the most important peculiarities of NIRCam astrometry, to make the reader aware of the subtleties that are present in the system.\\

Different levels of precisions may be needed for different use cases. For distant objects, e.g. TNOs, the general recommendations are often sufficient: an astrometric record which reports sufficiently conservative uncertainties, using the recommendations of this document, provides valuable astrometric information and is usually acceptable.\\

For nearby objects, such as NEOs, the faster motion and lower distance introduce effects that cannot be neglected. For these objects, astrometry should be submitted only when the specific aspects outlined in this guide have been taken into consideration, and properly corrected for. In these situations, we encourage submitters to interact with the SARC and with the JWST experts at STScI before finalizing and submitting their astrometric reports.

\appendix
\section{Appendix: Pipeline commands} \label{appendix}

To run the JWST pipeline with the {\tt clean\_flicker\_noise} step but without executing the {\tt jump} or the {\tt saturation} steps, set their {\tt skip} flags to {\tt True}, as in the following example:

\begin{verbatim}
from jwst.pipeline import Detector1Pipeline, Image2Pipeline

steps = {`jump': {`skip': True}, saturation: {`skip': True}, 
    `clean_flicker_noise': {`skip': False}}

Detector1Pipeline.call(<uncal file>, steps = steps, 
    save_results = True)

Image2Pipeline.call(<rate file>, save_results = True)
\end{verbatim}

\noindent The above example will result in Level 2b CAL and I2D files. All pipeline steps can be turned on or off individually; explicitly turning on default steps is not required. Additional details on the JWST pipeline can be found in the JWST documentation\footnote{\url{https://jwst-docs.stsci.edu/jwst-science-calibration-pipeline}} and on Read the Docs\footnote{\url{https://jwst-pipeline.readthedocs.io/en/latest/jwst/user\_documentation/introduction.html}}.

\end{document}